\documentclass[%
 reprint,
 amsmath,amssymb,
 aps,
]{revtex4-1}

\usepackage{graphicx}
\usepackage{dcolumn}
\usepackage{bm}
\usepackage{amssymb,amsmath}
\usepackage{braket}
\usepackage{epstopdf}

\begin{document}

\preprint{APS/123-QED}

\title{Measurement of the Kr\,{\small XVIII} {\it 3d} $^2$D$_{5/2}$ lifetime at low energy in a unitary Penning Trap}

\author{Nicholas D. Guise$^{1,2}$}
\email{Nicholas.Guise@gtri.gatech.edu}
\altaffiliation[present address: ]{Georgia Tech Research Institute, Atlanta, GA 30332}%
\author{Joseph N. Tan$^1$, Samuel M. Brewer$^{2,1}$}
\altaffiliation[Present address: ]{National Institute of Standards and Technology, Boulder, CO, 80305}
\author{Charlotte F. Fischer$^1$}
\author{Per J\"{o}nsson$^3$}
\address{$^1$ National Institute of Standards and Technology (NIST), 100 Bureau Drive,
Gaithersburg, Maryland 20899-8422, USA}
\address{$^2$ University of Maryland, College Park, Maryland 20742, USA}
\address{$^3$ Malm\"{o} University, S-20506, Malm\"{o}, Sweden}

\date{\today}
\begin{abstract}
A different technique is used to study the radiative decay of a metastable state in multiply ionized atoms. With use of a unitary Penning trap to selectively capture Kr$^{17+}$ ions from an ion source at NIST, the decay of the {\it 3d} $^2$D$_{5/2}$ metastable state is measured in isolation at low energy, without any active cooling.  The highly ionized atoms are trapped in the fine structure of the electronic ground configuration with an energy spread of 4(1) eV, which is narrower than within the ion source by a factor of about 100. By observing the visible 637 nm photon emission of the forbidden transition from the {\it 3d} $^2$D$_{5/2}$ level to the ground state, we measured its radiative lifetime to be $\tau = 24.48\,{\rm ms}\,\pm 0.28_{\text{stat.}}\,{\rm ms}\,\pm 0.14_{\text{syst.}}$ ms. Remarkably, various theoretical predictions for this relativistic Rydberg atom are in agreement with our measurement at the 1\% level. 

\end{abstract}


\maketitle


Forbidden transitions between long-lived states in atoms, which are metastable because they cannot decay via electric dipole transitions, play pivotal roles in many fields---including frequency-standard metrology \cite{Rosenband2010, Bloom2014, Derevianko2012}, determination of fundamental constants \cite{Hansch2013, CODATA2010}, tests of the standard model \cite{Hansch2013, Haxton2001}, and astrophysics \cite{Edlen1942, Mason1988}.  Isolating such systems at low temperature is ideal for understanding their characteristics and for developing applications. While laser techniques are useful in trapping and cooling neutral \cite{phillips1998laser} and singly ionized atoms \cite{ion_RMP2003}, the isolation and cooling of highly ionized atoms is, in general, made more challenging by the higher temperatures during production in ion sources. Highly charged ions have been successfully produced at low temperatures in ion traps via \textit{in situ} photoionization \cite{Church1987}, but this technique requires a synchrotron radiation source and produces a mixture of charge states.  Different types of ion traps have been used to capture and store single charge states of multiply ionized atoms \cite{Church1995,Gruber2001}, and evaporative cooling has been demonstrated recently \cite{Hobein2011}. However, the usefulness can be limited by the time required to cool the ions, which can exceed the time scale of interest---such as in radiative decay. For example, a time constant of $t_c \approx 32$ ms for evaporative cooling \cite{Hobein2011} would require 147 ms to reduce energy by a factor of 100.

In this work, we report the observation of a forbidden transition in highly ionized Kr$^{17+}$ atoms (or Kr {\scriptsize	XVIII} in spectroscopic notation) that are isolated at low energy. The ion source is the electron beam ion trap (EBIT) at NIST  \cite{Ratliff1998}. The low kinetic energy ($\approx 5$ eV) is attained within 1 ms after ion extraction, and is obtained by using a unitary Penning trap to selectively capture Kr$^{17+}$ ions extracted from the EBIT \cite{Brewer2013}. Although forbidden transitions have been studied within an EBIT, the mixture of charge states and the proximity of the hot electron gun filament can complicate lifetime measurements, in some cases limiting precision \cite{Trabert2001}. To illustrate the new technique, we provide an improved lifetime measurement for the {\it 3d} $^2$D$_{5/2}$ metastable state in Kr$^{17+}$, chosen for this demonstration because its unique atomic structure is potentially useful for a determination of the fine-structure constant (an important fundamental constant in physics and metrology) if the underlying theory were more refined.

\begin{figure}
\includegraphics[angle=0,width=\columnwidth]{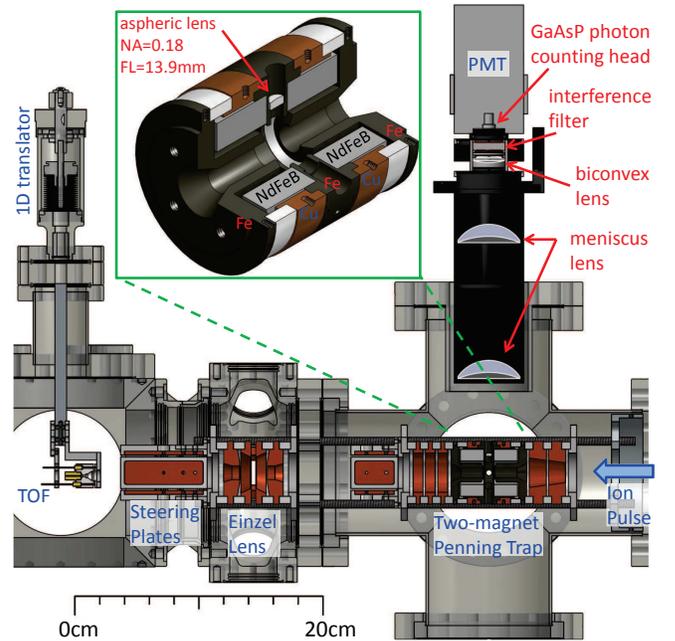}
\caption{\label{fig:KrSetup} Schematic diagram highlighting the unitary Penning trap (centered on the six-way cross, at right) used for capture and fluorescence detection of Kr$^{17+}$ ions. Ion pulses enter the apparatus from the right. Captured ions can be ejected to a time-of-flight (TOF) detector, at left.}
\end{figure}

Although potassium-like, the special properties of Kr$^{17+}$ are due to its deviation from the Madelung rule which normally assigns a {\it 4s} valence electron to the ground state of potassium (K) and singly ionized calcium (Ca$^+$).  Instead, the 19 electrons in Kr$^{17+}$ form a closed-shell [Ar] core with a {\it 3d} valence electron.  Interestingly, this high angular momentum {\it 3d} orbital is a Rydberg state (l=n-1).  However, relativistic effects are made rather conspicuous by the strong attraction of its screened nuclear charge ($Z'\,=\,Z-18\,=\,18$), which pulls the orbital closer to the nucleus and shifts the fine structure to the visible domain. We therefore use optical techniques to observe the forbidden decay of the upper {\it 3d} $^2$D$_{5/2}$ level via magnetic-dipole (M1) transition to the lower {\it 3d} $^2$D$_{3/2}$ level (ground state).

Figure \ref{fig:KrSetup} highlights the portion of the ion-capture apparatus that is used for light collection from Kr$^{17+}$ ions isolated in a unitary Penning trap \cite{Tan2012}. A full schematic-diagram and discussion of the ion capture process are provided in Ref. \cite{Brewer2013}.  Upon capture, ions of a selected charge state are radially confined by the 0.32-T field generated from two NdFeB magnets embedded within the soft-iron electrode assembly.  Along the trap axis, a confinement well is generated by higher electric potential applied on the endcap electrodes (annular) relative to the central ring.  

  Krypton gas is injected into the EBIT to produce multiply charged ions by electron-impact ionization; some of the Kr$^{17+}$ ions are collisionally excited to the metastable {\it 3d} $^2$D$_{5/2}$ level. A mixture of charge states is extracted from the EBIT and ejected in an ion pulse of width $<5$ $\mu$s .   An analyzing magnet in the extraction beamline selects out the Kr$^{17+}$ ions, yielding a narrower ion pulse of width $\approx 100$ ns; the selected ions are steered via electrostatic optics towards the ion capture apparatus (Fig.~\ref{fig:KrSetup}). The transit time over the $\approx 8$-m beamline from the EBIT to the Penning trap is $\approx$ 22 $\mu$s for Kr$^{17+}$ ions, much shorter than the radiative lifetime to be measured. The process of isolating ions in a unitary Penning trap \cite{Brewer2013} has enabled a selected charge state to be captured with an energy distribution of about 3 to 5 eV (roughly $100 \times$ lower than typically found in the EBIT ion source) without any active cooling scheme.  The lower energy contributes to reduction of systematic uncertainties in the lifetime measurement.
	
Kr$^{17+}$ ions in the {\it 3d} $^2$D$_{5/2}$ level emit 637-nm photons when they undergo magnetic dipole (M1) transitions to the {\it 3d} $^2$D$_{3/2}$ ground state. As shown in Fig.~\ref{fig:KrSetup}, the 637-nm fluorescence is collected with a lens system and detected by a photomultiplier tube (PMT) with a GaAsP photon-counting head (5 mm active diameter).  Mounted in one of the holes of the Penning trap ring electrode (see inset of Fig.~\ref{fig:KrSetup}), a small aspheric lens sits 13.2 mm above the trap center, with an effective numerical aperture of 0.18.  Along the vertical arm of the six-way cross, the collected light passes through three other lenses in air and an optical interference filter before reaching the PMT detector. The interference filter (center at $640\pm2$ nm) has a 10 nm bandpass to suppress stray light and cascade photons from charge-exchange products. The PMT counting head is cooled to $-20\, ^\circ\rm{C}$ using a Peltier device to obtain an {\it in situ} dark count rate of $\approx 5.5$ s$^{-1}$.    

\begin{figure}
\includegraphics[angle=0,width=0.45\textwidth]{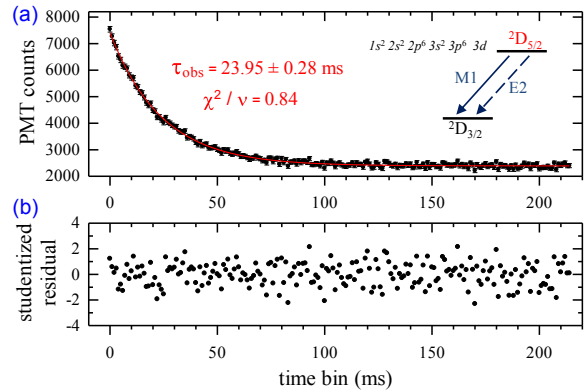}
\caption{\label{fig:LowPDecay} PMT signal from radiative (M1) decay of Kr {\scriptsize XVIII} ions captured in a unitary Penning trap at $1.0 \times 10^{-7}$ Pa ($8.0 \times 10^{-10}$ T) chamber pressure: (a) Photon counts in 1-ms time bins summed over repetitions in $\approx 26$ hours, with best fit (red curve) to a single-exponential function $N(t)=N_0 e^{-t/\tau_{\text{obs}}}+\eta$ . (b) Studentized residuals $[N(t)-N_i]/\sigma_i$ for the fit in panel (a).}
\end{figure}

Figure \ref{fig:LowPDecay} shows the fluorescence decay curve observed for the lowest chamber pressure.  Photons detected by the PMT are counted in 1-ms time bins by a fast multichannel scaler, synchronized to start at the instant of ion capture in the unitary Penning trap.  After light collection over several decay lifetimes, the captured ions are ejected from the trap, and a new excited population is loaded from the EBIT to start the next data acquisition cycle.  The combined data fit very well to a single exponential decay curve, $N(t)=N_0 e^{-t/\tau_{\text{obs}}}+\eta$, where the constant offset $\eta$ is due to the dark count of the PMT.  The goodness of fit is shown in the $\chi$-squared per degree of freedom [Fig.~\ref{fig:LowPDecay}(a)] and the distribution of studentized residuals [Fig.~\ref{fig:LowPDecay}(b)].  

\begin{table}
\caption{\label{tab:KrXVIIIhistory} Calculated and measured values for the lifetime $\tau$ of the {\it 3d} $^2$D$_{5/2}$ level in Kr {\scriptsize XVIII}.} 
\begin{ruledtabular}
\begin{tabular}{clll}
\textrm{\bf{Item}} & \bf{Source/method} & $\lambda$ (nm) &\textrm{\bf{$\tau$ (ms)}}\\
\hline
   & \multicolumn{2}{c}{Calculation:} &  \\
\cline{2-4}
1 &\cite{AliKim1988} Dirac-Fock\footnotemark[1]& 642.0 &  $24.57$ \\
  &                          & \footnotesize{Adjusted} & $24.02(5)$\footnotemark[3] \\
2 &\cite{Biemont1989} Relativistic Hartree-Fock & 636.82 & $23.85$ \\
  &                      & \footnotesize{Adjusted} & $23.89(5)$\footnotemark[3] \\
3 &\cite{Crespo1999} Cowan Code & 638.7 & $24.0$ \\
  &          & \footnotesize{Adjusted} & $23.83(5)$\footnotemark[3] \\
4 &\cite{Trabert2001} Relativistic Hartree-Fock & \hspace{0.15in}-- & $23.87$  \\
5 &\cite{Charro2002} Relativistic Quantum- & 626.2 & 22.78 \\
  & \hspace{0.50in}Defect Orbital    & \footnotesize{Adjusted} & $24.00(5)$\footnotemark[3] \\
6 & \cite{Ralchenkoprivcomm} Flexible Atomic Code    & \footnotesize{Adjusted} & $24.16(5)$\footnotemark[3]\\
7 &\cite{Sapirsteinprivcomm} Lowest-order RMBPT\footnotemark[2]& \footnotesize{Adjusted} & $24.02(5)$\footnotemark[3] \\
8 & \hspace{0.25in}GRASP2K \footnotesize{(this work)}& \footnotesize{Adjusted} & $24.02(5)$\footnotemark[3] \\
\cline{2-4}
   & \multicolumn{2}{c}{Measurement:}  & \\
\cline{2-4}
9 & \multicolumn{2}{l}{\cite{Trabert2001}\hspace{0.05in}Intra-EBIT experiment}  & $22.7 ~\pm 1.0$\\
10 &\multicolumn{2}{l}{\hspace{0.25in}Penning trap \footnotesize{(this work)}}   & $24.48 \pm 0.32$\\
\end{tabular}
\end{ruledtabular}
\footnotetext[1]{Single configuration approximation.} 
\footnotetext[2]{Relativistic many-body perturbation theory (RMBPT).}
\footnotetext[3]{Standard error based only on uncertainty of the semi-empirical Ritz wavelength $\lambda_\text{Ritz} \approx 637.2(4)$ nm from Ref. \cite{KAU89}.}

\end{table}

At the lowest chamber pressure ($P = 1.0 \times 10^{-7}$ Pa) the measured lifetime is $23.95\,{\rm ms} \pm 0.28$ ms, which agrees rather well with most of the calculations listed in Table \ref{tab:KrXVIIIhistory}. However, this measurement must be corrected slightly due to known systematic effects, for which a detailed account will be presented elsewhere. In brief, nonradiative processes that reduce the measured lifetime include collisional quenching of the metastable state, electron capture from background gas atoms, and ion orbit instability. Systematic effects are small for the lowest chamber pressure and the low-energy distribution of stored ions. The chamber pressure was controlled and varied to measure the pressure dependence of the decay rate; the resulting plot ($\tau^{-1}$ vs $P$, a Stern-Volmer plot) is used for extrapolation to the unperturbed $P=0$ limit. Moreover, from known Penning-trap dynamics and constraints imposed by TOF measurements, light-collection loss due to ion orbit instability (ions leaving the field of view of the first lens) can be estimated by simulation. Accounting for these small nonradiative losses, we determined that the {\it 3d} $^2$D$_{5/2}$ level has a radiative lifetime of $\tau = 24.48\,{\rm ms}\,\pm 0.28_{\text{stat.}}\,{\rm ms}\,\pm 0.14_{\text{syst.}}$ ms. The pressure offset gives the main systematic uncertainty, due to gauge calibration and residual gas composition.  

\begin{table*}
\caption{ {\sc grasp2k} computations for the M1 transition rate, $A(M1)=\tau^{-1}$, using different active orbital sets for optimization. The adjusted values (last column) are obtained by rescaling with the semi-empirical $\lambda_\text{Ritz} = 637.186\, {\rm nm}$ (see Ref.\cite{KAU89}).}
\label{tab:grasp2k} 
\begin{ruledtabular}
\begin{tabular*}{180mm}{@{} c c l c c c  @{}}
   & \multicolumn{2}{c}{Wave function expansion strategy} & $\,\lambda^{\text{pred.}}\,$ & \,$A^{\text{pred.}}$ \, &  \,$A^{\text{adj.}}$ \,\\
Index & Single and double excitations from & Subshell with 1 excitation & (${\rm nm}$) & (s$^{-1}$)\, &  (s$^{-1}$)\,\\
\hline
1 & \footnotesize{$3s^23p^63d$}\hspace{1.4in} & \footnotesize{ Required: $3s^2$ and/or $3p^6$}& 645.703 & 40.00 & 41.62 \\
2 & \footnotesize{$3s^23p^63d$, \,$3s^23p^43d^3$, \,and $3s3p^63d^2$} & \footnotesize{ No constraint}  & 642.343 & 40.63 & 41.62 \\
3 & \footnotesize{$3s^23p^63d$, \,$3s^23p^43d^3$, \,and $3s3p^63d^2$} & \footnotesize{ Allowed: $2s^2$ or $2p^6$}\hspace{0.20in} & 640.328 & 41.02 & 41.63 \\
\end{tabular*}
\end{ruledtabular}
\end{table*}

\begin{figure}
\includegraphics[angle=0,width=0.49\textwidth]{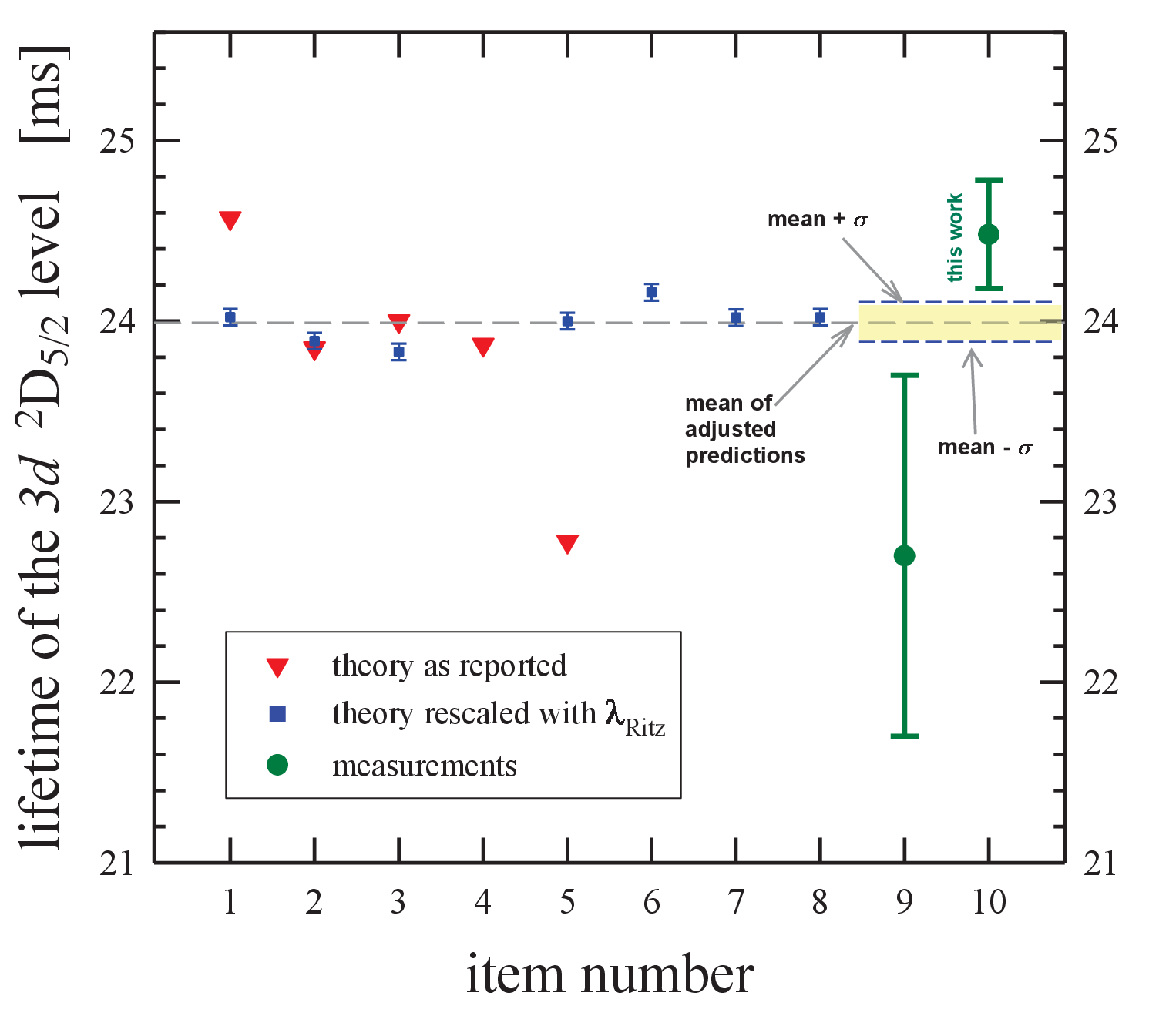}
\caption{\label{fig:KrHistory} Lifetime of the Kr {\scriptsize XVIII} {\it 3d} $^2$D$_{5/2}$ level: Comparison of measurements (circle) with theoretical values, as reported (triangle) or adjusted (square). Items plotted are listed in Table \ref{tab:KrXVIIIhistory}.}
\end{figure}

Measurements are compared with various theoretical works in Fig. \ref{fig:KrHistory}.  Table \ref{tab:KrXVIIIhistory} lists the experimental results and theoretical predictions. The $3d\ ^2D_{5/2}$ level decays mainly by the spin-flipping M1 transition; the rate for electric quadrupole (E2) transition is negligible \cite{AliKim1988}. Relativistic calculations prior to this work (tabulated as items 1-5 of Table \ref{tab:KrXVIIIhistory}) are depicted as triangles in Fig. 3, showing a considerable spread (with deviation as large as 5$\sigma$ from this work) due to the uncertainty in the transition wavelength.  However, using the more precise wavelength compiled in the NIST database, the predictions from different methods can be rescaled and brought into agreement at the 1\% level.  The rescaling is straightforward since, using Fermi's golden rule, the M1 transition rate is formally similar to the expression for the electric dipole transition, and therefore has the familiar inverse cubic dependence on the transition wavelength: 
 
\begin{equation}
\label{eq:Fermi}
A = \frac{1}{\tau} = \frac{(2\pi)^2}{h}\sum_f{\rho(E_f)|\bra{f} V_{int} \ket{i}|^2} \propto \frac{1}{\lambda^3} \;,
\end{equation}

\noindent where $h$ is the Planck constant, $V_{int}$ is the interaction potential for the transition from the initial state $\ket{i}$ to the final state $\bra{f}$, and $\rho(E_f)$ is the density of final states per unit energy. Wavelength-adjusted theoretical results are included in Table \ref{tab:KrXVIIIhistory} and shown as squares in Fig. \ref{fig:KrHistory}; the reported uncertainty is propagated from the semiempirical Ritz wavelength used for the rescaling.

The results from three additional calculations are reported here (items 6-8 of Table \ref{tab:KrXVIIIhistory}), representing diverse computational methods.  The mean of all wavelength-adjusted lifetime predictions is 23.99 ms; remarkably, the standard deviation of 0.11 ms is $< 0.5$\%, depicted in Fig. \ref{fig:KrHistory} as horizontal dashed lines.  This mean is slightly lower than our measurement after corrections, a mild discrepancy of $< 1.5 \,\sigma_\text{combined}$.

It is atypical to find such close convergence of results from a diverse array of methods---including lowest-order perturbation theory, single-configuration approximations, as well as sophisticated variational optimization involving multiple subshells and excitations. We examined this convergence with calculations using GRASP2K, a general-purpose atomic structure package (version 2K) that is suitable for large scale relativistic calculations based on multi-configuration Dirac-Hartree-Fock theory \cite{Jonsson2007}. Table \ref{tab:grasp2k} provides the results of three GRASP2K calculations, showing improvement of {\it ab initio} predictions (columns $\lambda^{\text{pred}}$ and $A^{\text{pred}}$) by refining the set of active electron orbitals generated for the wave function expansion. We also give the adjusted transition rate (last column) obtained using the more precise Ritz wavelength in Ref. \cite{KAU89}. We find that the adjusted transition rate is remarkably stable, regardless of the complexity of configuration mixing; that is, the transition matrix in Eq. (\ref{eq:Fermi}) is insensitive to configuration interaction. This feature can be useful for more precise tests to examine the higher order effects from quantum electrodynamics (QED), which has come under scrutiny due to the discrepancy in proton size measurements \cite{Pohl2010}.

The measurement with a unitary Penning trap is $1.78$\,ms longer than the previous measurement inside an EBIT \cite{Trabert2001}---a mild discrepancy of 1.7 $\sigma_{\rm{combined}}$.  This work has three times less uncertainty than the previous measurement, partly because thermal radiation from the EBIT electron gun did not add significantly to the background noise for this method, as it did for the intra-EBIT measurement  \cite{Trabert2001}. But also noteworthy is the reduction of systematic effects enabled by low-energy ions under better control in a unitary Penning trap.

In summary, forbidden transitions in highly charged ions can be studied precisely in a unitary Penning trap.  A unique feature is that a single charge state of interest is isolated at relatively low energy ($\approx 5$ eV) within 1 ms after capture \cite{Brewer2013}. Thus, in many cases, highly charged ions prepared in a metastable state during production can be studied at low energy, on a time scale that is too short for efficient application of known cooling techniques.  By using the finer control in this well-characterized system, an improved measurement of the Kr {\scriptsize XVIII} {\it 3d} $^2$D$_{5/2}$ lifetime is obtained at the 1\% level of precision.  In addition to its forbidden transition being accessible to optical frequency combs, this unusual K-like atom is of particular interest because its valence electron is in a Rydberg state ({\it 3d} orbital).  The high angular momentum simplifies theoretical considerations. Nevertheless, it seems fortuitous to find that diverse theoretical calculations of the {\it 3d} $^2$D$_{5/2}$ lifetime converge at the one-percent level.  This augurs well for higher accuracy. For example, relativistic many-body perturbation theory (RMBPT) could perhaps be refined to provide {\it ab initio}, high precision predictions---which, in conjunction with further experimental progress, could lead to a measurement of the fine-structure constant.  Experimentally, a more precise measurement of the fine-structure splitting could improve comparisons of theoretical predictions, especially when higher-order corrections are included.  Finally, the technique demonstrated here can be useful for studying a variety of unexplored metastable states---such as those proposed recently for enhancing sensitivity to Lorentz symmetry breaking from possible time variation of the fine-structure constant \cite{Berengut2010}, or for developing new, ultrastable atomic clocks \cite{Derevianko2012}.

\begin{acknowledgments}
The work of N.D.G. was supported in part by a National Research Council Research Associateship Award at NIST.  We thank P.J. Mohr, J. Sapirstein, and Yu. Ralchenko for stimulating discussions.
\end{acknowledgments}

%

\end{document}